\documentclass[journal,10pt,twoside,a4paper]{IEEEtranTCOM}
\usepackage[utf8]{inputenc}
\usepackage[english]{babel}
\usepackage[T1]{fontenc}
\usepackage{color}
\usepackage{graphicx}
\usepackage{array}
\makeatletter

\newbox\@eqnb@x\setbox\@eqnb@x\hbox{}
\def\@@eqncr{\let\reserved@a\relax
    \ifcase\@eqcnt \def\reserved@a{& & &}\or \def\reserved@a{& &}%
     \or \def\reserved@a{&}\else
       \let\reserved@a\@empty
       \@latex@error{Too many columns in eqnarray environment}\@ehc\fi
     \reserved@a \if@eqnsw\@eqnnum\stepcounter{equation}\fi\@eqnnumm
     \global\@eqnswtrue\global\@eqcnt\z@\cr}
\def\@eqnnumm{\hb@xt@.01\p@{}%
        \rlap{\hskip-\displaywidth\box\@eqnb@x}}
\def\intertextsl#1{\global\setbox\@eqnb@x\hbox{\hskip\leftskip%
                \normalfont #1}}
\makeatletter
\def\ve#1{{\mathchoice{\mbox{\boldmath$\displaystyle #1$}}%
              {\mbox{\boldmath$\textstyle #1$}}%
              {\mbox{\boldmath$\scriptstyle #1$}}%
              {\mbox{\boldmath$\scriptscriptstyle #1$}}}}
\def\tfrac#1#2{{\textstyle\frac{#1}{#2}}}
\def\Pr{\mathop{\mathrm{Pr}}\nolimits}

\def\mod{\mathop{\mathrm{mod}}}

\def\erf{\mathop{\mathrm{erf}}\nolimits}

\def\e{\mathrm{e}}
\def\d{\mathrm{d}}
\def\T{\mathsf{T}}

\def\dB{\mathrm{dB}}
\DeclareSymbolFont{EULE}{U}{eur}{m}{n}
\DeclareMathAlphabet{\matheule}{U}{eur}{m}{n}
\DeclareMathSymbol{\pdf}{\mathalpha}{EULE}{"66}
\def\defeq{\stackrel{\smash{\mbox{\scriptsize def}}}{=}}
\DeclareSymbolFont{EULE}{U}{eur}{m}{n}
\DeclareMathSymbol{\bn}{\mathalpha}{EULE}{"30}
\DeclareMathSymbol{\be}{\mathalpha}{EULE}{"31}
\DeclareMathSymbol{\dirac}{\mathalpha}{EULE}{"0E}
\DeclareMathAlphabet{\mathfrak}{U}{euf}{m}{n}
\SetMathAlphabet{\mathfrak}{bold}{U}{euf}{b}{n}
\DeclareMathAlphabet{\ff}{U}{euf}{m}{n}
\SetMathAlphabet{\ff}{bold}{U}{euf}{b}{n}
\DeclareSymbolFont{EULE}{U}{eur}{m}{n}
\DeclareMathSymbol{\pdf}{\mathalpha}{EULE}{"66}
\DeclareSymbolFont{AMSb}{U}{msb}{m}{n}
\DeclareMathSymbol{\C}{\mathalpha}{AMSb}{"43}
\DeclareMathSymbol{\F}{\mathalpha}{AMSb}{"46}
\DeclareMathSymbol{\R}{\mathalpha}{AMSb}{"52}
\DeclareMathSymbol{\N}{\mathalpha}{AMSb}{"4E}
\DeclareMathSymbol{\Z}{\mathalpha}{AMSb}{"5A}
\DeclareMathAlphabet{\rv}{OT1}{lmss}{m}{sl}
\DeclareMathAlphabet{\rve}{OT1}{lmss}{bx}{sl}
\DeclareMathAlphabet{\mathfrak}{U}{euf}{m}{n}
\SetMathAlphabet{\mathfrak}{bold}{U}{euf}{b}{n}
\DeclareMathAlphabet{\ff}{U}{euf}{m}{n}
\SetMathAlphabet{\ff}{bold}{U}{euf}{b}{n}
\DeclareSymbolFont{AMSa}{U}{msa}{m}{n}
\DeclareMathSymbol{\sm}{\mathbin}{AMSa}{"39}
\def\mumo{\mu\sm 1}
\graphicspath{{./}}

\begin{document}
\title{Helper Data Schemes for Coded Modulation and Shaping in
	Physical Unclonable Functions}
\IEEEoverridecommandlockouts
\author{Robert F.H. Fischer%
\thanks{Robert F.H. Fischer is with the
	Institute of Communications Engineering, Ulm University,
	89081 Ulm, Germany
	(e-mail: robert.fischer@uni-ulm.de).
}
}
\markboth{ }{ }
\maketitle

\begin{abstract}
In this paper, we consider the generation and utilization of helper data for
physical unclonable functions (PUFs) that provide real-valued readout symbols.
Compared to classical binary PUFs, more entropy can be extracted from each
basic building block (PUF node), resulting in longer keys/fingerprints and/or
a higher reliability. To this end, a coded modulation and signal shaping
scheme that matches the (approximately) Gaussian distribution of the readout
has to be employed.
A new helper data scheme is proposed that works with any type of coded
modulation/shaping scheme. Compared to the permutation scheme from the
literature, less amount of helper data has to be generated and a higher
reliability is achieved.
Moreover, the recently proposed idea of a two-metric helper data scheme is
generalized to coded modulation and a general S-metric scheme. It is shown
how extra helper data can be generated to improve decodability.
The proposed schemes are assessed by numerical simulations and by evaluation
of measurement data. We compare multi-level codes using a new rate design
strategy with bit-interleaved coded modulation and trellis shaping with a
distribution matcher. By selecting a suitable design, the rate per PUF
node that can be reliably extracted can be as high as 2~bit/node.
\end{abstract}

%
%

\section{Introduction}
\label{sec_1}

\noindent
The extraction of a unique fingerprint from integrated circuits is a field of
current research and enables various applications ranging from key generation
and authentication to the identification of (e.g., safety-critical)
components. Due to uncontrollable variations in the manufacturing process of
microelectronic devices, the randomness of so-called Physically Unclonable
Functions (PUFs) is unique, uncontrollable, and non-reproducible, see,
e.g., \cite{maes2013physically}.

PUFs are composed of \emph{PUF nodes},%
\footnote{Other names used in the literature are \emph{PUF cell}, cf., e.g.,
        \cite{Mueelich:21,Park:21} (not to be confused with ``memory
        cell'') or \emph{PUF unit}, cf., e.g., \cite{Zhang:21}.
}
each of which delivers a single random variable, see, e.g.,
\cite{immler2019new}. There are several basic principles for constructing
a PUF node, e.g., ring oscillator PUFs, arbiter PUFs, or those based on the
power-on state of memory cells. By combining $n$ (independent) PUF nodes,
the PUF is obtained.
We consider so-called ``weak'' PUFs, where upon request a readout word of
length $n$ (the result from the $n$ nodes) is delivered.%
\footnote{The coined denomination ``\emph{function}'' is somewhat misleading
	in the context of weak PUFs. Upon a trigger, an (almost) fixed
	\emph{readout} is provided. Given this readout, the final
	\emph{response} or \emph{key} is derived.
}
In contrast, in ``strong'' PUFs, the response depends on a challenge.

The exploited (desired) randomness that makes a PUF instance unique occurs
in the manufacturing process; then the PUF can be assumed to be static over
its lifetime.
However, repeatedly extracted readouts may vary (slightly) due to a change,
e.g., in temperature or supply voltage, or due to ageing effects.
This unwanted randomness must be counteracted by channel coding to obtain
a unique and stable fingerprint.

The vast majority of the literature deals with PUFs that deliver a \emph{binary
readout}, e.g.,
\cite{maes2012pufky,mueelich2014error,puchinger2015error,Mueelich:21},
and the respective binary (hard-decision) channel coding approaches.
However, the readout is extracted from an analog source. Using the analog
(real-valued) readout, or at least the reliability information extracted
from the PUF node, significant reliability improvements can be achieved, e.g.,
\cite{muelich2019channel,maes2009soft,maes2009low,yu2010secure,taniguchi2013stable,Mueelich:21}.

Besides increasing the reliability, more than a single binary symbol (bit)
may be generated from the analog readout, since an analog (continuous-valued)
source contains more than one bit of entropy. Various approaches to so-called
\emph{multi-valued PUFs} exist, e.g., (the list is not exhaustive)
\cite{tuyls2006read,bossuet2013puf,gunlu2014dct,willers2016mems,kodytek2016temperature,chuang2017physically,chuang2018multi,immler2018b,immler2019secure,immler2019higher,immler2019variable,mandry2020normalization,pehl2020spatial,kazumori2019ternary,kazumori2020debiasing}.

In contrast to the schemes employing multilevel quantization, in
\cite{Mueelich:21,Fischer:22}, we have presented approaches that directly
use the the analog, non-quantized output of a PUF node. Interpreting the
readout process in PUFs as a digital transmission scheme, \emph{coded
modulation} and \emph{signal shaping} are designed and applied. This allows
longer fingerprints/keys to be generated from a given number of PUF nodes
with high reliability.

In conventional digital transmission, the transmitter is guaranteed to
generate a valid codeword. In PUFs, the readout will most likely not be a
valid codeword. In order to enable the application of channel coding/coded
modulation, the PUF readout must be brought into the form of a valid codeword
with superimposed error. To this end, a \emph{helper data scheme} is used.
The most prominent approach to this task in hard-decision binary schemes is
the code-offset algorithm \cite{juels1999fuzzy,linnartz2003new,dodis2004fuzzy}.
In \cite{Fischer:22}, a first helper data scheme was presented that works
with coded modulation/shaping.

\begin{figure*}
\centerline{\includegraphics[width=.725\textwidth,clip=]{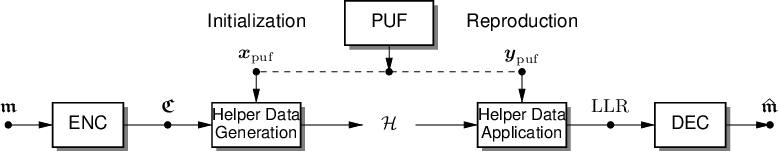}}
\caption{\label{fig_1_1}
	The concept of helper data generation in the initialization phase
	(left part) and the usage of this helper data in the reproduction
	phase (right part).
}
\end{figure*}

Helper data is generated and used as follows.
In the \emph{initialization} of the \emph{enrollment}, which takes place in
a secure environment right after the PUF has been manufactured,
the actual readout $\ve{x}_\mathrm{puf}$ is measured and set as the
\emph{reference} (or nominal) readout.
In addition, a \emph{message} word $\ve{\ff{m}}$ is randomly selected and
encoded by the used coded modulation scheme (for mathematical details and
notation see Sec.~\ref{sec_2}).
Based on the codeword and the reference PUF readout, the helper data is
generated, see Fig.~\ref{fig_1_1}.
At \emph{reproduction}, when the fingerprint is requested, the (noisy) PUF
readout $\ve{y}_\mathrm{puf}$ is extracted. Based on this word and using the
helper data, a channel decoder is able to decode the message $\ve{\ff{m}}$.

In this interpretation, the message $\ve{\ff{m}}$ establishes the fingerprint
or key. The PUF serves as a source of (noisy) common randomness and the
transmission takes place via the helper data. This point of view shows that
the helper data plays an important role. First of all, the helper enables
decoding. However, if (additional) helper data is generated in a suitable way,
decoding can be improved.

The present paper deals with the generation of helper data for coded modulation
and shaping. Messages $\ve{\ff{m}}$ with the maximum possible length should
be recovered as reliably as possible in the reconstruction phase.

The contribution of the present paper is as follows:

i) We propose a new helper data scheme that works for any type of coded
modulation and/or signal shaping scheme. Compared to the permutation scheme
of \cite{Fischer:22} less amount of helper data has to be generated. However,
at the same time a higher reliability is achieved. To this end, a refined
model of the PUF as digital transmission with randomness at the transmitter
is discussed.

ii) The idea of the two-metric helper data scheme introduced in
\cite{Danger:19,Tebelmann:21} is generalized to coded modulation and a general
$S$-metric scheme. It is shown how extra helper data can be generated which
improves the decodability.

iii) In order to optimize the performance of the used code with relatively
short block length, we propose a new rate design strategy for multi-level
codes. This approach is based on the actual performance of the component
codes rather than the capacities of the bit levels only.

The schemes are presented in detail and evaluated using numerical simulations
and measurement data from PUF realizations. Please note that the paper focuses
on the PUF model, the helper data generation, and the application of coding and
shaping. It is shown that the proposed helper data does not reveal any
information about the secret message. However, active attacks on the PUF
are beyond the scope of this paper.

The paper is organized as follows: Sec.~\ref{sec_2} gives preliminary remarks
and discusses a refined model of the assumed soft-output PUF where the two
main random effects (manufacturing and readout process) are clearly separated.
In Sec.~\ref{sec_3} the new helper data scheme is presented and compared to
the existing one. Its security is proven and it is shown how optimal decoding
can be carried out.
The $S$-metric scheme is discussed in Sec.~\ref{sec_4}. The generation of
the extra helper data is explained and it is shown that it works for both
uniform and shaped signaling. Optimum and possible suboptimum decoding is
discussed as well as the asymptotic performance. 
Finally, in Sec.~\ref{sec_5}, results from numerical simulations are compiled
covering the trade-off between the amount of helper data generated and
reliability.
The paper is concluded in Sec.~\ref{sec_6}.
A convenient transformation of Gaussian random variables, which facilitates
the presentation, is given in Appendix~\ref{sec_a}.

%
%

\section{PUF Model and Preliminaries}
\label{sec_2}

\noindent
In this section, we discuss a model for PUFs that deliver real-valued random
variables. The model is suitable for assessing and designing coded
modulation/shaping schemes.

\subsection{Statistic of the PUF Readout}

A well-suited assumption is that the reference PUF readout
$x_\mathrm{puf}$ is zero-mean Gaussian distributed; w.l.o.g.\ it can be
normalized to unit variance. The readout of a PUF node at reproduction is
$y_\mathrm{puf} = x_\mathrm{puf} + e_\mathrm{puf}$.
The error $e_\mathrm{puf}$ occurring in repeated readouts can be expected to
be independent of the reference readout and also to be a zero-mean Gaussian
random variable with some variance $\sigma_\rv{e}^2$.
This model has been justified by an exhaustive measurement campaign at the
Institute of Microelectronics, Ulm University, for ROPUFs,
see \cite{herkle2019depth,Mueelich:21}.

When dealing with PUFs, two types of randomness are present. The one at the
manufacturing process, which delivers $x_\mathrm{puf}$ (from then on fixed for
the PUF node), and which is the desired randomness. The other is the unwanted
randomness $e_\mathrm{puf}$ at reproduction, which we assume to be drawn
independently at each readout.

The PUF is composed of $n$ PUF nodes. A common assumption is that the PUF nodes
are independent of each other and all have the same statistics, i.e., they are
i.i.d. (memory effects as in \cite{Maes:13} are not considered here).
The PUF readout is thus given by the vector%
\footnote{Notation:
	We distinguish between scalars (normal font) and vectors (bold font).
	Here, all vectors are row vectors.
	We also distinguish between quantities from the set of real numbers $\R$
	(conventional italics) and variables over the binary field $\F_2$
	(Fraktur font). Random variables are typeset in sans-serif font.
}
\begin{eqnarray}						\label{eq_2_1}
\ve{y}_\mathrm{puf} &=& \ve{x}_\mathrm{puf} + \ve{e}_\mathrm{puf} \;,
\end{eqnarray}
which has i.i.d.\ Gaussian components.

\subsection{Interpretation as Digital Transmission}

For the design of coded modulation schemes, it is rewarding to interpret the
PUF and its associated randomness as a digital transmission scheme, cf.\ 
\cite{Mueelich:21,Fischer:22}, i.e., that information has to be conveyed from a
transmitter to a receiver. To this end, in order to ensure reliable reception,
a codeword is generated from the information to be communicated, utilizing
some channel code. W.l.o.g.\ we restrict ourselves to binary component
codes.

In $M$-ary signaling ($M = 2^\mu$), each element of the codeword of length
$n$ is represented by a binary $\mu$-tuple, the \emph{label}
$\ve{\ff{c}}_i = [\ff{c}_{\mumo,i}\; \ldots\; \ff{c}_{0,i}]$, $i=1,\ldots,n$.
Combining these $\mu$-tuples column-wise the \emph{codematrix}
\begin{eqnarray}
\ve{\ff{C}} &=& [\ve{\ff{c}}_1^\T\; \ldots\; \ve{\ff{c}}_n^\T]
\\
		&=& \left[ \matrix{\ff{c}_{\mumo,1} & \ff{c}_{\mumo,2}
				& \cdots & \ff{c}_{\mumo,n} \cr
				\vdots & \vdots & & \vdots \cr
				\ff{c}_{0,1} & \ff{c}_{0,2}
					& \cdots & \ff{c}_{0,n}} \right]
		\;=\; \left[ \matrix{\ve{\ff{c}}^{(\mumo)} \cr
				\vdots \cr \ve{\ff{c}}^{(0)}} \right] \quad
\end{eqnarray}
is obtained; its rows are denoted by $\ve{\ff{c}}^{(m)}$, $m=0,\ldots,\mu-1$.

In classical digital communications, the labels $\ve{\ff{c}}_i$ are eventually
mapped to unique real (or complex) numbers $a_i$, called \emph{signal points}.
The set of all possible signal points is the \emph{signal constellation}.

When dealing with soft-output PUFs the way of thinking is different.
Instead of signal points, $M$ \emph{regions} are defined \cite{Fischer:22}.
The set of regions $\mathcal{R}_\rho$, $\rho = 0,\ldots,M-1$, where $\rho$
is the region number, constitutes a partition of the real line, i.e.,
\begin{eqnarray}						\label{eq_2_10}
\bigcup_{i=0}^{M-1} \mathcal{R}_{\rho_i} \;=\; \R \;, \quad && \quad
	\mathcal{R}_{\rho_i} \cap \mathcal{R}_{\rho_j} \;=\; \{\}
					\;,\quad \forall i \neq j \;.
\end{eqnarray}
W.l.o.g.\ we number the regions according to their position on the real line.
Equivalently, we may characterize the regions by their lower and upper limits
(we assume that the regions are compact), i.e.,
\begin{eqnarray}						\label{eq_2_11}
\mathcal{R}_\rho &=& [\, L_\rho,\; L_{\rho+1}\, ) \;,
\\
\intertextsl{with}
-\infty = L_0 < L_1 < &\ldots& < L_{M-1} < L_M = +\infty \;.
\end{eqnarray}

Additionally, a one-to-one mapping from a binary $\mu$-tuple
$\ve{\ff{c}} = [\ff{c}_{\mumo}\ldots\ff{c}_{1}\ff{c}_{0}]$ (the columns of
$\ve{\ff{C}}$) to a region number $\rho$ has to be defined
\begin{eqnarray}						\label{eq_2_12}
\mathcal{M}: \F_2^\mu \mapsto \{0,\;1,\ldots,\;M-1\} \;,\qquad
			\rho = \mathcal{M}( \ve{\ff{c}} ) \;.
\end{eqnarray}
Given the mapping, region number $\rho$ and binary label
$\ve{\ff{c}} = [\ff{c}_{\mumo}\,\ldots\,\ff{c}_{1}\,\ff{c}_{0}]$ are equivalent
and subsequently used interchangeably whatever denomination is more suited,
e.g., $\mathcal{R}_\rho$, $\mathcal{R}_\ve{\ff{c}}$, and
$\mathcal{R}_{[\ff{c}_{\mumo}\ldots\ff{c}_{1}\ff{c}_{0}]}$ mean the same.

Basically, two mappings are of interest. On the one hand,
\emph{natural labeling} which is identical to the one-dimensional
\emph{set-partition labeling}. Here, the region number is given by simply
reading the label as binary number $\rho = [\ff{c}_{\mumo}\cdots \ff{c}_0]_2$,
where $\ff{c}_0$ is the least significant bit (LSB).
On the other hand, \emph{Gray labeling} (in particular binary reflected Gray
labeling), where the bit labels of adjacent regions differ in a single bit
position, can be used.

Moreover, given the regions $\mathcal{R}_\rho$, a \emph{quantizer}
\begin{eqnarray}						\label{eq_2_13}
\mathcal{Q}: \R \mapsto \F_2^\mu \;, && \quad
			\ve{\ff{q}} = \mathcal{Q}( x )
\end{eqnarray}
can be defined. For a real-valued number $x$, it returns the binary $\mu$-tuple
$\ve{\ff{q}}$ corresponding to the region number where $x$ lies.

\begin{figure}[ht]
\centerline{\includegraphics[width=.95\columnwidth,clip=]{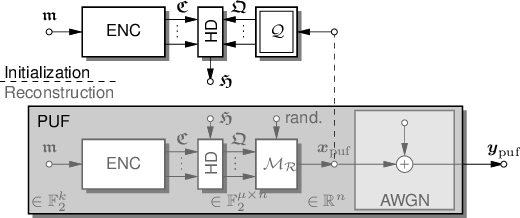}}
\caption{\label{fig_2_1}
	Model of the PUF as communication system with $M$-ary ($M = 2^\mu$)
	signaling and readout process modeled as transmission over an AWGN
	channel.
	Upper part: operations in the initialization phase.;
	Lower part: interpretation for the readout process in the
	reconstruction phase.
}
\end{figure}

Putting all ingredients together, the PUF readout process can be modeled as
a digital transmission scheme, see Fig.~\ref{fig_2_1}. Please note that for
characterizing the readout process we  \emph{imagine} the lower part;
the upper part gives the operations which are actually carried out in the
initialization phase.

In the initialization phase, the reference PUF readout
$\ve{x}_\mathrm{puf} = [x_{\mathrm{puf},1}, \ldots, x_{\mathrm{puf},n}]$
is determined. The manufacturing process is the respective random experiment.
The reference $\ve{x}_\mathrm{puf}$ is then fix for each PUF instance.
The quantizer (\ref{eq_2_13}) is used to determine the labels $\ve{\ff{q}}_i$
of the regions where $x_{\mathrm{puf},i}$, $i=1,\ldots,n$, lies
(upper part of Fig.~\ref{fig_2_1}). The labels are written column-wise into
the matrix $\ve{\ff{Q}}$ (with the LSB at the bottom row).

In the reconstruction phase, the interpretation is as follows. The label
$\ve{\ff{q}}_i$ indicates the region where $x_{\mathrm{puf},i}$ lies. The
actual number is modeled to be drawn according to the portion of a Gaussian
distribution over the respective region. All points within a region represent
the same codesymbol. This randomness is part of the manufacturing process, but
in terms of digital communications, it is the randomness at the transmitter
(mapping $\mathcal{M}_\mathcal{R}$ with additional randomness).
The manufacturing randomness is thus conceptually split into two parts:
the randomness that defines the regions used to carry information, and the
randomness within the region that is not utilized (for the moment).

As mentioned above, the readout process at reconstruction, which results in
$\ve{y}_\mathrm{puf}$, is modeled as the transmission of $\ve{x}_\mathrm{puf}$
over an AWGN channel with noise variance $\sigma_\rv{e}^2$ per component.

Finally, the generation of the region labels $\ve{\ff{q}}_i$ at reproduction
needs to be modeled. To this end, we again take a look at the initialization
phase. As in classical communications, a binary message word of length $k$,
i.e., $\ve{\ff{m}} \in \F_2^k$, is drawn at random. This is another source of
randomness, as the word $\ve{\ff{m}}$ is drawn at random during the
initialization process. Given a coded modulation/shaping scheme (e.g.,
those in \cite{Fischer:22}), the code matrix $\ve{\ff{C}}$ is generated from
$\ve{\ff{m}}$. This matrix has to be linked to the matrix $\ve{\ff{Q}}$ of
region labels which is present for the particular PUF instance.
This is accomplished by the \emph{helper data (HD) scheme}---given
$\ve{\ff{C}}$ and $\ve{\ff{Q}}$, helper data $\ve{\ff{h}}$ is generated.

In the reconstruction phase, we \emph{imagine} that the message word
$\ve{\ff{m}}$ is encoded by the coded modulation/shaping scheme.
Applying the helper data, the region labels are obtained and fed to the
mapper as discussed above.

It is worth noting that this model generalizes the classical code-offset
algorithm \cite{juels1999fuzzy,linnartz2003new,dodis2004fuzzy} in two aspects.
On the one hand, the real-valued output is considered by introducing the
(random) mapping and by replacing the binary symmetric channel by the AWGN
channel.
On the other hand, instead of binary transmission, an $M$-ary, $M>2$, scheme
is present. Thus, not a single binary codeword is considered, but a code
matrix with $\log_2(M)$ rows. This generalization also calls for suitable
helper schemes.

%
%

\section{Helper Data Schemes for Coded Modulation and Shaping}
\label{sec_3}

\noindent
We now turn to helper data schemes, which are suited for coded modulation and
shaping schemes. We have to distinguish two approaches for the selection of
the regions $\mathcal{R}_\rho$, which essentially determine the properties
of the coding scheme.

\subsection{Regions, Distributions, and Shaping}

As we expect the PUF readout $\rv{x}$ to be zero-mean Gaussian with variance
$\sigma_\rv{x}^2 = 1$, the regions are selected with probability
\begin{eqnarray}						\label{eq_3_1}
p_\ve{\ff{q}} &=& \int_{\mathcal{R}_\ve{\ff{q}}}
			\frac{1}{\sqrt{2\pi}}\,\e^{-\frac{x^2}{2}} \;\d x \;
	= \int_{L_{\rho}}^{L_{\rho+1}}
			\frac{1}{\sqrt{2\pi}}\,\e^{-\frac{x^2}{2}} \;\d x \;, \;
\end{eqnarray}
where $\rho = \mathcal{M}(\ve{\ff{q}})$ is the region number.
The (conditional) pdf of the  PUF readout when using this region is
\begin{eqnarray}						\label{eq_3_2}
\pdf_\rv{x}(x \mid \ve{\ff{q}})
	\;=\; &=& \cases{\frac{1}{p_\ve{\ff{q}}}
				\frac{1}{\sqrt{2\pi}}\, \e^{-\frac{x^2}{2}},
			& $x \in \mathcal{R}_\ve{\ff{q}}$ \cr
		     0, & else} \;.
\end{eqnarray}

A first approach to defining the regions is to choose the region limits $L_\rho$
in such a way that $p_\ve{\ff{q}} = \frac{1}{M}$, $\forall \rho$, cf., e.g.,
\cite{immler2019new,mandry2020normalization}.
Thus, \emph{uniform signaling} is present. The coded modulation scheme
has to generate the region labels $\ve{\ff{q}}_i$ such that they are uniformly
distributed.
Using the transformation introduced in the Appendix~\ref{sec_a}, the
limits for an $M$-ary uniform scheme are given by
$L_\rho = g^{-1}(\frac{2}{M}\,\rho -1)$,
$\rho = 0,\ldots,M$ (with $L_0 = -\infty$ and $L_M = \infty$).

In contrast, the (inner) limits $L_\rho$, $\rho = 1,\ldots,M-1$, may be spaced
uniformly and symmetrically around the origin ($L_{M/2} = 0$), cf., e.g.,
\cite{Fischer:22}. This gives rise to \emph{shaped signaling}. The coded
modulation/shaping scheme has to be designed so that the probabilities of the
region labels $\ve{\ff{q}}_i$ follow the distribution (\ref{eq_3_1}).
In this case, the width $\lambda$ of the (inner) regions is a free parameter
which has to be optimized.

\begin{figure}[ht]
\centerline{\includegraphics[width=.70\columnwidth,clip=]{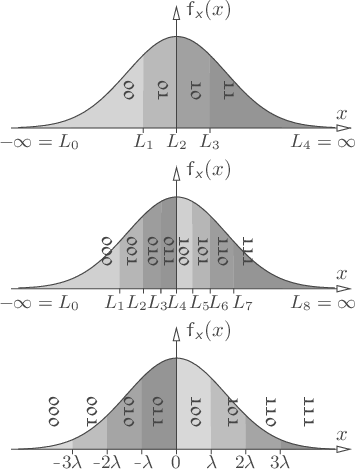}}
\caption{\label{fig_3_1}
	Regions $4$-ary and $8$-ary uniform signaling,
	$8$-ary shaped signaling (top to bottom). Natural labeling.
}
\end{figure}

Fig.~\ref{fig_3_1} visualizes the regions and the respective portions of the
Gaussian distribution for (from top to bottom) $4$-ary and $8$-ary uniform
signaling and $8$-ary shaped signaling.

%

\subsection{Helper Data Schemes}
\label{sec_3B}

Conventional digital transmission ensures that the transmitter generates a
valid codeword. In case of PUFs, the readout will most likely not be a valid
codeword/codematrix. Consequently, a so-called \emph{helper data scheme}
is employed to transform the PUF readout to a valid codeword---in signal space,
to a valid sequence of amplitudes and signs.

\subsubsection{Binary Case}

In binary hard-decision PUFs, the helper data is obtained by an element-wise
addition over $\F_2$ (XOR) of the binary reference readout
$\ve{\ff{x}}_\mathrm{puf}$ with the binary codeword $\ve{\ff{c}}$, i.e.,
$\ve{\ff{h}} = \ve{\ff{x}}_\mathrm{puf} \oplus \ve{\ff{c}}$
\cite{juels1999fuzzy,linnartz2003new,dodis2004fuzzy}.
When binary codes with soft-decision decoding are employed, a sign flip is
able to establish a valid codeword from an arbitrary readout \cite{Mueelich:21}.
Therefore, in the binary case, a single bit of helper data is required per
binary codesymbol.

\subsubsection{Coded Modulation---Permutation}

\begin{figure}[ht]
\centerline{\includegraphics[width=.98\columnwidth,clip=]{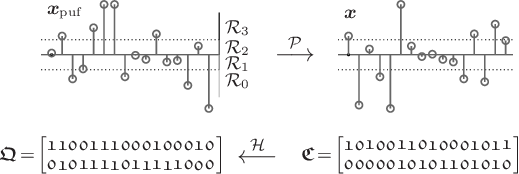}}
\caption{\label{fig_3_2}
	Example for helper schemes.
	The PUF reference readout $\ve{x}_\mathrm{puf}$ (top left) is given.
	The elements of the word in the signal space should lie in the regions
	indicated by the columns of the codematrix $\ve{\ff{C}}$ (bottom right).
	A permutation and possibly a sign flip of the elements
	leads to the word $\ve{x}$ (top right), which matches the demands
	except for $x_1$.
	The proposed helper scheme indicates which bits of $\ve{\ff{C}}$
	have to be flipped in order to obtain the matrix $\ve{\ff{Q}}$
	(bottom left), whose columns indicate in which regions the elements
	of $\ve{x}_\mathrm{puf}$ lie.
}
\end{figure}

In \cite{Fischer:22}, a first helper data scheme for PUFs employing coded
modulation/shaping has been presented. Since in $M$-ary signaling, a simple
sign flip is not sufficient, a signed permutation has been proposed.
The permutation and sign flip are chosen in such a way that the processed PUF
readout w.r.t.\ amplitude and sign matches the desired codeword in signal space.
This approach is visualized in the upper part of Fig.~\ref{fig_3_2}.
It works for both uniform and shaped signaling.

This scheme has two drawbacks. First, due to the statistics of the reference
readout, it is not guaranteed that there is a perfect match between the
desired codeword and the permuted/inverted readout, in the sense that all
samples are in the regions indicated by the code symbols.
Second, the number of bits required to represent the helper data is upper
bounded by $n\,(1 + \log_2(n))$ ($n$ for the sign flips and
$\log_2(n!) = \sum_{l=1}^n \log_2(l) < n\log_2(n)$ for the permutation).

%
%
%

\subsubsection{Coded Modulation/Uniform---Conversion}

We now pre\-sent a new helper data scheme for coded
modulation/\discretionary{}{}{}shaping that circumvents both aforementioned
drawbacks. For the moment, we assume uniform signaling.

The permutation is obtained by answering the question, which modifications
have to be applied to the reference PUF readout $\ve{x}_\mathrm{puf}$ in the
signal space in order to obtain (approximately) a valid codeword.

Reverse thinking leads to a better approach. The codematrix $\ve{\ff{C}}$ is
given by the selected message word $\ve{\ff{m}}$ encoded with the chosen
coded modulation/shaping scheme. The reference PUF readout $\ve{x}_\mathrm{puf}$
is quantized and thus characterized by the binary words $\ve{\ff{q}}_i$
specifying the regions or, in total, by the matrix $\ve{\ff{Q}}$, cf.\
Fig.~\ref{fig_2_1}.

The helper data has to indicate which bits of the desired code matrix
$\ve{\ff{C}}$ need to be flipped in order to get the given matrix
$\ve{\ff{Q}}$. This is simply obtained as
\begin{eqnarray}						\label{eq_3_10}
\ve{\ff{H}} &=& \ve{\ff{Q}} \oplus \ve{\ff{C}} \;.
\end{eqnarray}
We call this \emph{conversion scheme}; it is a generalization of the
code-offset scheme. Two $\mu \times n$ matrices are XORed instead of two words
of length $n$.

First, the security of this approach is proven. Subsequently, we show how
optimal decoding is conducted.

A helper data scheme has to fulfill three demands. First, knowing the PUF
readout $\ve{y}_\mathrm{puf}$ (or even the noise-free reference readout
$\ve{x}_\mathrm{puf}$) and the helper data $\ve{\ff{H}}$, the message
$\ve{\ff{m}}$ has to be decodable. Second, if only the PUF readout is known,
no leakage about the message $\ve{\ff{m}}$ must occur. Third, if only the
helper data is known, no leakage about the message $\ve{\ff{m}}$ must occur.
The leakage is quantified by the mutual information $\mathrm{I}(\cdot;\cdot)$
between the quantities of interest.

\paragraph{Decodability}
Suppose that the PUF readout $\ve{x}_\mathrm{puf}$ (noise-free case) and the
helper data $\ve{\ff{H}}$ are known. Given a hypothesis $\ve{\tilde{\ff{c}}}$
for the label, the region label number
$\ve{\tilde{\ff{q}}} = \ve{\tilde{\ff{c}}} \oplus \ve{\ff{h}}$
can be calculated. Using this converted label and the PUF readout, the decoding
metric (see below) can be derived. In the noise-free case, the correct
hypothesis will have a probability of one; all other will have a probability
of zero. Thus, $\ve{\ff{C}}$
is known and since $\ve{\ff{C}} = \mathsf{ENC}(\ve{\ff{m}})$ is a one-to-one
function of the message, $\ve{\ff{m}}$ can be recovered.
In the noisy case, there is no perfect knowledge; with some probability,
a decoding error will occur.

\paragraph{No Leakage when Knowing the PUF Readout Only}
Since the message $\ve{\ff{m}}$ is drawn independently of the reference
readout $\ve{x}_\mathrm{puf}$, by definition we have
\begin{eqnarray}						\label{eq_3_12}
\mathrm{I}(\ve{\ff{m}};\; \ve{x}_\mathrm{puf}) &=& 0 \;;
\end{eqnarray}
no information about $\ve{\ff{m}}$ can be extracted if only
$\ve{x}_\mathrm{puf}$ is known.

\paragraph{No Leakage when Knowing the Helper Data Only}
Finally, assume that the helper data matrix $\ve{\ff{H}}$ is known.
Due to the assumption of independent PUF nodes, the columns $\ve{\ff{q}}_i$
of the matrix $\ve{\ff{Q}}$ of the quantized reference readout are independent.
In the case of uniform signaling, the vectors
$\ve{\ff{q}}_i = [\ff{q}_{\mu\sm 1,i}\ldots\ff{q}_{0,i}]$ are uniformly
distributed. As a consequence the elements $\ff{q}_{m,i}$,
$m=0,\ldots,\mu\sm 1$, $i=1,\dots,n$, are uniform and independent of each
other. The helper data $\ff{h}_{m,i} = \ff{q}_{m,i} \oplus \ff{c}_{m,i}$
is thus independent of $\ff{c}_{m,i}$ ($\ve{\ff{Q}}$ acts as as one-time pad
for $\ve{\ff{C}}$). As $\ve{\ff{m}}$ and $\ve{\ff{C}}$ are related one-to-one
by the encoding procedure, we finally have
\begin{eqnarray}						\label{eq_3_13}
\mathrm{I}(\ve{\ff{m}};\; \ve{\ff{H}}) &=& 0 \;.
\end{eqnarray}

\subsubsection{Coded Modulation/Shaping---Conversion}

The situation changes for schemes employing signal shaping. Here the
region labels $\ve{\ff{c}}_i$ and $\ve{\ff{q}}_i$ have a non-uniform
distribution (the probabilities are given by the areas of the Gaussian density
in Fig.~\ref{fig_3_1} within the regions). Consequently, $\ve{\ff{q}}_i$ is
not a perfect one-time pad for $\ve{\ff{q}}_i$ (and vice versa), and knowing
$\ve{\ff{h}}_i = \ve{\ff{q}}_i \oplus \ve{\ff{c}}_i$ provides some knowledge
about $\ve{\ff{c}}_i$ and, finally, about the message $\ve{\ff{m}}$.

A simple modification solves this problem. We expect the readouts of the
PUF nodes (index $i$) to be independent. Thus, a certain element of
$\ve{\ff{q}}_j = [\ff{q}_{\mu\sm 1,j}\ldots\ff{q}_{0,j}]$ is independent of all
elements of $\ve{\ff{c}}_i = [\ff{c}_{\mu\sm 1,i}\ldots\ff{c}_{0,i}]$ for
$i \neq j$. Only if all entries of $\ve{\ff{q}}_j$ are treated jointly and
are combined with all elements of $\ve{\ff{c}}_i$, a leakage would occur.
If the $\mu$ bits $\ff{q}_{m,j}$ all come from different positions $j$ this
can be avoided.%
\footnote{To be precise, the \emph{symbol-by-symbol} leakage is avoided.
	When looking at the entire matrix
	$\ve{\ff{H}} = [\ve{\ff{h}}_1^\T \ldots \ve{\ff{h}}_n^\T]$, the
	dependencies are still present. However, these can only be exploited
	if the entire codematrix $\ve{\ff{C}}$ is inferred from the entire
	matrix $\ve{\ff{H}}$. Compared to randomly guessing the codematrix
	(and thus the message $\ve{\ff{m}}$), there is only a marginal
	advantage in practice.
}
In summary, calculating the helper data according to
\begin{eqnarray}						\label{eq_3_15}
\ff{h}_{m,i} &=& \ff{q}_{m,\mod_n(i+m\cdot o-1)+1} \oplus \ff{c}_{m,i} \;,
		\;\; {i=1,\dots,n \atop m=0,\ldots,\mu\sm 1} \;,\quad
\end{eqnarray}
where $o \neq 0$ is some fixed offset and $\mod_n(\cdot)$ the usual modulo
operation, no (relevant) leakage occurs.

\subsection{LLR Calculation}

We now turn to the calculation of the decoding metric, specifically
\emph{log-likelihood ratios (LLR)}, for the proposed helper data scheme.
Note that the noisy PUF readout $\ve{y}_\mathrm{puf}$ is given by
(\ref{eq_2_1}).

We are interested in the LLR for the label bit $\ff{c}_{m,i}$ (level $m$,
position $i$ within the codeword). For this the conditional pdf of the PUF
output is required, assuming that this label bit has the given value and
knowing the helper data $\ve{\ff{h}}_i$.
Using the result from \cite{Fischer:22}, it can be written as%
\footnote{Please note that a factor $\frac{1}{p_\ve{\ff{c}}}$ is missing
	in Equation $(17)$ in \cite{Fischer:22}.
}
\begin{eqnarray}                                                \label{eq_3_30}
&& \hskip-8mm\pdf_\rv{y}(y_{\mathrm{puf},i} \mid \ff{c}_{m,i},\, \ff{h}_{m,i})
\\
&=& \pdf_\rv{y}(y_{\mathrm{puf},i} \mid
			\ff{q}_{m,i} = \ff{c}_{m,i} \oplus \ff{h}_{m,i} )
\nonumber\\
&=& \frac{1}{p_{\ff{q}_m}} \, \frac{1}{\sqrt{2\pi\sigma_\rv{e}^2}}\,
			\e^{-\frac{y^2_{\mathrm{puf},i}}{2(1+\sigma_\rv{e}^2)}}
	\!\!\!\!\!\!
	\sum_{\forall \ve{\ff{q}}, \ff{q}_m = \ff{c}_{m,i} \oplus \ff{h}_{m,i}}
	\!\!\!\!\!\!
	\Delta\mathcal{Q}(y_{\mathrm{puf},i}, \mathcal{R}_\ve{\ff{q}} ) \;,
\nonumber
\end{eqnarray}
where the following abbreviation has been used
\begin{eqnarray}						\label{eq_3_31}
\Delta\mathcal{Q}(y, \mathcal{R}_{\ve{\ff{c}}} )
	&=&  \mathcal{Q}( D\,L_{\rho} - F\,y )
		- \mathcal{Q}( D\,L_{\rho+1} - F\,y ) \;, \quad
\end{eqnarray}
with $D \defeq \sqrt{(1 + \sigma_\rv{e}^2)/\sigma_\rv{e}^2}$, \
$F \defeq 1/\sqrt{(1 + \sigma_\rv{e}^2)\sigma_\rv{e}^2}$, and
$\mathcal{Q}(x) \defeq \int\nolimits_{x}^{\infty} \frac{1}{\sqrt{2\pi}}\,
			\e^{-\frac{z^2}{2}} \; \d z$ is the
\emph{complementary Gaussian integral function}. $L_{\rho}$ and $L_{\rho+1}$ are
the limits of the regions $\mathcal{R}_\ve{\ff{q}}$ (cf.\ (\ref{eq_2_11})).

The LLR is then given as
\begin{eqnarray}						\label{eq_3_32}
\mathrm{LLR}(\ff{c}_{m,i}) &=&
		\log\left( \frac{\Pr\{ \ff{c}_{m,i} = \ff{0} \mid y \}}
                                {\Pr\{ \ff{c}_{m,i} = \ff{1} \mid y \}} \right)
\nonumber\\
&=& \log\left( \frac{\pdf_\rv{y}(y_{\mathrm{puf},i} \mid
				\ff{q}_{m,i} = \ff{0} \oplus \ff{h}_{m,i} )
				p_{\ff{q}_m = \ff{0}} }
                    {\pdf_\rv{y}(y_{\mathrm{puf},i} \mid
				\ff{q}_{m,i} = \ff{1} \oplus \ff{h}_{m,i} )
				p_{\ff{q}_m = \ff{1}} }
			\right)
\nonumber\\
&=& \log\left(
        \frac{\sum_{\forall \ve{\ff{q}}, \ \ff{q}_m = \ff{0} \oplus \ff{h}_{m,i}}
		\Delta\mathcal{Q}(y_{\mathrm{puf},i}, \mathcal{R}_\ve{\ff{q}} )}
	     {\sum_{\forall \ve{\ff{q}}, \ \ff{q}_m = \ff{1} \oplus \ff{h}_{m,i}}
		\Delta\mathcal{Q}(y_{\mathrm{puf},i}, \mathcal{R}_\ve{\ff{q}} )}
			\right) \;.\quad
\end{eqnarray}
In the same way, the LLR for a label bit given (having already decoded) some
other label bit(s) can be stated, see \cite{Fischer:22}. There, the summation
runs over all regions where the known label bit has the given value.
In the case of shaping, the shift in (\ref{eq_3_15}) between $\ff{q}_{m,j}$ and
$\ff{c}_{m,i}$ has to be additionally taken into account.

%
%

\section{$S$-Metric Helper Data Scheme}
\label{sec_4}

\noindent
In the first place, the helper data enables decoding to take place at all.
However, decoding can be improved if (additional) helper data is generated
in a suitable way. In \cite{Danger:19,Tebelmann:21}, a two-metric helper data
scheme has been proposed for a binary readout per PUF node. In this section,
we generalize this idea to $M$-ary coded modulation and $S$-metric schemes.

\subsection{Regions and Helper Data}

The main idea of the approach in \cite{Danger:19} is to produce helper data 
in the initialization phase that do not only guarantee that decoding is
possible in principle, but also contain some form of reliability information
about the PUF readout. Specifically, a binary variable is generated that
indicates (in the case of hard decision) which of two possible quantizers
should be used in the reconstruction.
To this end, the two quantization cells are each divided into two finer cells.
The reference PUF readout is classified in which of the finer cells it lies;
this establishes the helper data (in case of channel coding the extra helper
data over that used in the code-offset scheme). In the reconstruction phase
this extra data assists the decoding. As usual, it has to be guaranteed
that the helper data does not reveal any information about the PUF readout.

The idea of \cite{Danger:19,Tebelmann:21} can be generalized to $M$-ary
approaches and to $S$-metric schemes as follows. The exposition employs the
``tilde domain'' introduced in Appendix~\ref{sec_a} (all quantities are
marked with a tilde), where the actual Gaussian distribution is transformed
by $g(x)$ onto a uniform one over the interval $[-1,\; +1]$.

For an $S$-metric scheme, each of the $M$ regions $\mathcal{R}_\rho$ is
subdivided into $S$ subregions $\mathcal{R}_{\rho,s}$, $\rho=0,\ldots,M-1$,
$s=0,\ldots,S$. The limits of the subregions are uniformly spaced in the
tilde domain. Let the limits $L_\rho$ of the $M$-ary approach be given and
let $\tilde{L}_\rho = g(L_\rho)$. The limits for an $M$-ary $S$-metric
scheme are then specified by
\begin{eqnarray}						\label{eq_4_1}
\tilde{L}_{\rho,s} &=& \tilde{L}_\rho
		+ \frac{\tilde{L}_{\rho+1} - \tilde{L}_\rho}{S}\;s
		\;,\qquad {\rho=0,\ldots,M-1 \atop s=0,\ldots,S-1} \;.\quad
\end{eqnarray}
By construction,
\begin{eqnarray}						\label{eq_4_2}
p_{\rho,s} &=& \Pr\{ x \in \mathcal{R}_{\rho,s} \}
	= \Pr\{ x \in \mathcal{R}_\rho \} \cdot \frac{1}{S} \;.
\end{eqnarray}
Regardless of the initial limits (uniform or shaped signaling),
the PUF readout lies with equal probabilities ($\frac{1}{S}$) in the
subregions and region number $\rho$ and subregion number $s$ are independent.

The procedure in the initialization phase is as follows. The reference
$\ve{x}_\mathrm{puf}$ is read out from the PUF.
The vector's elements, $x_{\mathrm{puf},i}$, are quantized using a quantizer
that employs the limits $L_{\rho,s}$ (in the original domain). This gives
i) the region number $\rho_i = 0,\ldots,M-1$, and ii) the number of the
subregion $s_i=0,\ldots,S-1$.
The region number $\rho_i$ is used to generate the first part of the helper
data, namely the matrix $\ve{\ff{H}}$.
The subregion numbers, $s_i$, constitute the additional part of the helper
data. It may be expressed as a $\lceil \log_2(S) \rceil$ bit number
$s_i = [\;\ve{\ff{s}}_i\,]_2$ ($s_i$ and $\ve{\ff{s}}_i$ are used synonymously).
In summary, the total helper data is
\begin{eqnarray}						\label{eq_4_3}
\mathcal{H} &=& \{\, \ve{\ff{H}},\, \ve{\ff{S}} \,\} 
\end{eqnarray}
with $\ve{\ff{S}} = [\ve{\ff{s}}_1^\T,\ldots,\ve{\ff{s}}_n^\T]$.

The additional amount of helper data compared to the classical case (which is
obtained for $S=1$) is $n\cdot\log_2(S)$ bits.
The total amount of helper data for an $M$-ary $S$-metric scheme
is thus $n\,\big( \log_2(M) + \log_2(S) \big) = n\cdot\log_2(M\,S)$ bits.

\subsection{Security / Leakage}

This augmented helper data scheme fulfills the three demands discussed in
Sec.~\ref{sec_3B}. Decodability is guaranteed even without the additional
part of the helper data. There is still no leakage when only the PUF readout
is known. Since region number $\rho$ and subregion number $s$ are independent,
this additional part of the helper data does not reveal any information about
the codeword $\ve{\ff{c}}$ and thus the message $\ve{\ff{m}}$.
The subregion constitutes a new dimension that is orthogonal to the regions
which represent the codeword.

\subsection{Active Constellation and Decoding}

\begin{figure}[ht]
\centerline{\includegraphics[width=.98\columnwidth,clip=]{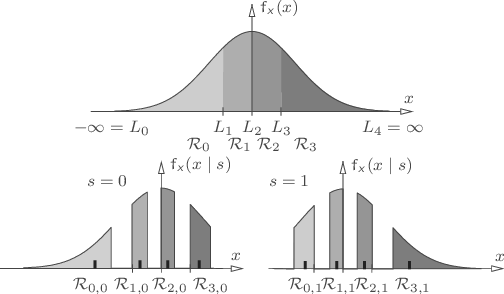}}
\caption{\label{fig_4_1}
	Regions and pdf the readout.
	Top: conventional case ($M=4$, uniform signaling);
	Bottom: $S=2$ and subregions/pdfs for $s=0$ and $s=1$.
	The centroid of the subregions are indicated by the black ticks.
}
\end{figure}

The two parts of the helper data ($\ve{\ff{H}}$ and $\ve{\ff{S}}$) have
different tasks. The matrix $\ve{\ff{H}}$ enable decoding in the first place.
The matrix $\ve{\ff{S}}$ improves decoding performance by providing additional
information about the readout. It is known from which subregion the reference
readout originates from. This is visualized in Fig.~\ref{fig_4_1}.

Conventionally, only the regions $\mathcal{R}_\rho$ are considered, which fill
the entire real line. As the PUF readout is continuous (and Gaussian), there
is no space between the regions. Having determined the subregions in the
initialization phase, and knowing the subregion number $s_i$ at the decoder,
the intervals from which $x_{\mathrm{puf},i}$ originate are restricted to the
subregions $\mathcal{R}_{\rho,s_i}$---a clear separation is visible.
Hence, the randomness at the transmitter in the model in Fig.~\ref{fig_2_1},
which is conventionally not known, is reduced, i.e., partially made available
to the receiver.
As $S$ increases, the subregions get smaller and parts of the density
concentrate more and more tending towards a discrete constellation.

The optimal decoding is done as described above. In the LLR calculation
(\ref{eq_3_32}), the regions $\mathcal{R}_{\ve{\ff{c}}}$ are simply replaced
by the subregions $\mathcal{R}_{\ve{\ff{c}},\ve{\ff{s}}_i}$. As a consequence,
the decoding process has the same complexity no matter which $S$ is chosen.
However, the exact metric calculation requires the repeated evaluation of
complementary Gaussian integral functions. Even for moderate $S$, the active
parts of the transmit density can be approximated by their centroids (which
are indicated in Fig.~\ref{fig_4_1}). Decoding may be done as for discrete
signal constellations transmitted over an AWGN channel.

\subsubsection{Asymptotic Performance}

As $S$ increases, the parts of the pdf $\pdf_\rv{x}(x \mid s)$ of $x$ given $s$
concentrate into narrow pulses and the pdf tends to be discrete.
In the tilde domain, the $M$ parts are uniformly spaced, leading to
non-uniformly spaced signal points in the original domain.
For $M$-ary signaling and $S \to \infty$, the signal points are
\begin{eqnarray}						\label{eq_4_10}
a_m(\varsigma) &=& g^{-1}\Big( \tfrac{2}{M}(m+\varsigma) - 1 \Big)
		\;, \quad {m=0,\ldots,M-1 \atop \varsigma \in [0,\; 1]}
		\;. \quad
\end{eqnarray}

Within each codeword various constellations with different distances between
the signal points are present. Since the receiver is aware of the present
constellation, the effect is similar to transmitting over a fading channel.
The asymptotic performance is obtained by averaging the performance of the
different constellations.

%
%

\section{Numerical and Experimental Results}
\label{sec_5}

\noindent
We now present results from numerical simulations. The main focus is on the
word error ratio (WER), which is the probability that the decoding result
$\ve{\ff{\hat{m}}}$ at reproduction differs from the message $\ve{\ff{m}}$
drawn at initialization. A WER below $10^{-6}$ is typically desired
\cite{Mueelich:21}.

\subsection{Setting and Parameters}

In principle, all coded modulation and shaping schemes may be used in the
present setting. Here we restrict ourselves to \emph{multilevel coding} (MLC)
\cite{Imai:77,Wachsmann:99} in combination with \emph{multistage decoding}
(MSD), and \emph{bit-interleaved coded modulation} (BICM)
\cite{Caire:98}. In case of shaping, \emph{trellis shaping}
\cite{Forney:92,Fischer:02} is used. Details on encoder and decoder structures
can be found in \cite{Fischer:22}. For comparison, the approach of
\cite{Boecherer:15}, which uses BICM in combination with a so-called
\emph{distribution matcher} (DM), is also considered. Here, the shaping is
realized by source decoding, cf.\ \cite{Fischer:02}.

Compared to the state of the art in \cite{Fischer:22}, the schemes are improved
in three aspects. First, the newly proposed helper data scheme is employed
to replace the permutation approach. Second, the Polar codes
\cite{arikan2009channel}, which are again employed, are designed differently.
For rates larger than $1/2$ the design of the frozen set based on the
Bhattacharyya parameter \cite{arikan2009channel,vangala2015comparative} is
still used. For lower rates, the sets are selected according to a technique
called $\beta$-expansion \cite{He:17}. Numerical simulations revealed some
advantage of this strategy for low-rate codes which are required in the
present setting.

\begin{figure}[htb]
\centerline{\includegraphics[width=.98\columnwidth,clip=]{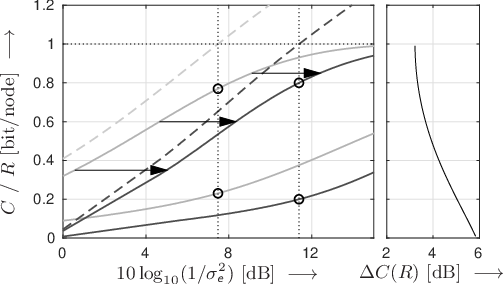}}
\caption{\label{fig_5_1}
	Left: Level capacities (solid) and sum capacities (dashed) over the
	signal-to-noise ratio (in dB) for a $4$-ary uniform scheme.
	Light gray: Capacities as derived in \cite{Fischer:22} and respective
	rate design (dotted line) for sum rate $R = 1.0$.
	Dark Gray: Capacity curves distorted by $\Delta C(R)$ (indicated by
	the arrows) and respective rate design.
	Right: Approximated distance (in dB) from the capacity limit
	$\Delta C(R)$ that results in the offset of the curves in the
	left figure.
}
\end{figure}

Third, a revised rate design for the multilevel codes is utilized.
In \cite{Fischer:22}, the rate design follows the capacity rule
\cite{Wachsmann:99} (see the details there). However, the Polar codes operate
at some distance from capacity, which, even more importantly, depends on the
rate. Low-rate codes perform significantly worse than codes with higher rates.
By numerical simulations of Polar codes (codelength $n = 1024$, various code
rates, BPSK over the AWGN channel), the required signal-to-noise ratio for
$\mathrm{WER} = 10^{-6}$ has been determined. The distance (in dB) from the
capacity limits is approximated by
$\Delta C(R) = \exp(-.27(1-R)^4 + .87(1-R)^2 + 1.17)$ via a polynomial fit.
The bit-level capacity curves derived in \cite{Fischer:22} are distorted by
$\Delta C(R)$ (for each rate $R$, the curve is shifted over the signal-to-noise
ratio by the respective amount). Based on these curves, the rates of the
component codes are selected. The procedure is visualized in Fig.~\ref{fig_5_1}.
The rates of the lower levels are decreased while the rates of the upper
levels are increased.

%

Subsequently, we compare the $4$-ary uniform, $8$-ary uniform, and $8$-ary
shaping (with $\lambda = 0.70$, cf.\ \cite{Fischer:22}) schemes.
Without further notice, we consider $n=1024$ PUF nodes and a target rate of
$R = 1.50$ resulting in the message length $k=1536$.
The Polar codes are decoded employing the standard \emph{successive
cancellation decoder} \cite{arikan2009channel} (list decoding \cite{Tal:15}
is not utilized).

The rates of the component codes when employing MLC (which have a codelength
$n=1024$) are collected in Tab.~\ref{tab_5_1}.
\begin{table}
\caption{\label{tab_5_1}
	Designs (coding rates $R_i$ of the component codes and code dimensions
	$k_i$ for the considered codelength $n = 1024$) used in the numerical
	simulations.
}
\centerline{%
\def\arraystretch{1.4}
\begin{tabular}{p{20mm}|>{\centering\arraybackslash}p{16mm}>{\centering\arraybackslash}p{16mm}>{\centering\arraybackslash}p{16mm}}
\hline
$R = 1.50$	& Level $0$	& Level $1$	& Level $2$
\\\hline\hline
$4$-ary uniform	& $R_0 = .511$  \newline  $k_0 =  523$ 
		& $R_1 = .989$  \newline  $k_1 = 1013$
		& \raisebox{-.4\baselineskip}{\hbox{---}}
\\\hline
$8$-ary uniform
		& $R_0 = .103$  \newline  $k_0 =  106$
		& $R_1 = .429$  \newline  $k_1 =  439$
		& $R_2 = .968$  \newline  $k_2 =  991$
\\\hline
$8$-ary shaping
		& $R_0 = .098$  \newline  $k_0 =  100$
		& $R_1 = .902$  \newline  $k_1 =  924$
		& $R_2 = .500$  \newline  $k_2 =  512$
\\\hline
\end{tabular}%
}
\end{table}

%

\subsection{Helper Schemes: Permutation vs.\ Conversion}

\begin{figure}[htb]
\centerline{\includegraphics[width=.90\columnwidth,clip=]{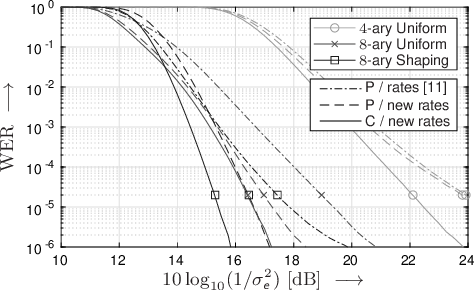}}
\caption{\label{fig_5_2}
	Word error ratio over the signal-to-noise ratio (in dB) for the
	$4$-ary uniform (light gray), $8$-ary uniform (middle gray),
	and $8$-ary (dark gray) shaping schemes with
	multilevel codes ($\lambda = 0.70$).
	$n=1024$ PUF nodes and $R = 1.50~\mathrm{bit/node}$.
	Dash-dotted: results from \cite{Fischer:22} (permutation (P) helper
	data scheme and rate design based on capacities).
	Dashed: permutation helper data scheme with new rate design taking
	the actual performance of the Polar codes into account.
	Solid: conversion (C) scheme from Sec.~\ref{sec_3} and new rate
	design.
}
\end{figure}

First, the permutation helper data scheme of \cite{Fischer:22} is compared
with the new scheme proposed in Sec.~\ref{sec_3}.
The word error ratios for the three schemes that employ multilevel codes
and multistage decoding are plotted in Fig.~\ref{fig_5_2}.

As can be seen, the permutation helper data scheme together with the rate
design based solely on capacities (the results from \cite{Fischer:22})
show the poorest performance. The $8$-ary schemes demonstrate clear
improvements when using the new rate design that considers the actual
performance of the Polar codes. The rates of the $4$-ary scheme are almost
the same for both strategies, as is the performance. By replacing the
permutation helper data scheme with the proposed conversion scheme, additional
gains can be achieved. For $\mathrm{WER} = 10^{-6}$ the $8$-ary shaping scheme
requires a signal-to-noise ratio less than $16~\dB$.

%

\subsection{S-Metric Helper Scheme}

Second, we consider the $S$-metric scheme of Sec.~\ref{sec_4}. Only the 
conversion scheme will be used for the one part of the helper data
($\ve{\ff{H}}$ in (\ref{eq_4_3})).
All other parameters remain the same as before ($n = 1024$,
$R = 1.50~\mathrm{bit/node}$, multilevel codes). 

\begin{figure}[htb]
\centerline{\includegraphics[width=.90\columnwidth,clip=]{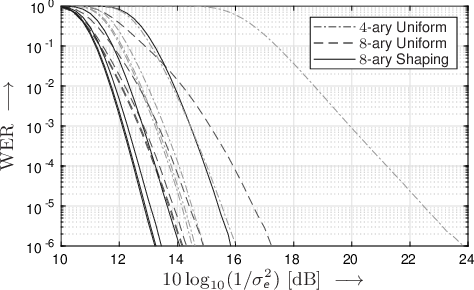}}
\caption{\label{fig_5_3}
	Word error ratio over the signal-to-noise ratio (in dB) for the
	$4$-ary uniform, $8$-ary uniform, and $8$-ary shaping schemes with
	multilevel codes ($\lambda = 0.70$).
	Conversion helper scheme from Sec.~\ref{sec_3} and new rate design.
	$S$-metric scheme. Right to left: $S = 1$, $2$, $4$, $8$, and $16$.
}
\end{figure}

In Fig.~\ref{fig_5_3} the word error ratios are plotted for $S = 1$, $2$,
$4$, $8$, and $16$ (right to left curves). Significant gains can be achieved for
$S > 1$ (most pronounced for the $4$-ary scheme). For $S > 8$ almost no extra
gain is provided. Looking at the $8$-ary shaping scheme, the performance
(at $\mathrm{WER} = 10^{-6}$) improves by approximately $1.5~\dB$ for $S=8$.
Nevertheless, the shaping scheme still outperforms the other variants.

%

\subsection{Helper Data vs.\ Performance Tradeoff}

\begin{figure}[htb]
\centerline{\includegraphics[width=.90\columnwidth,clip=]{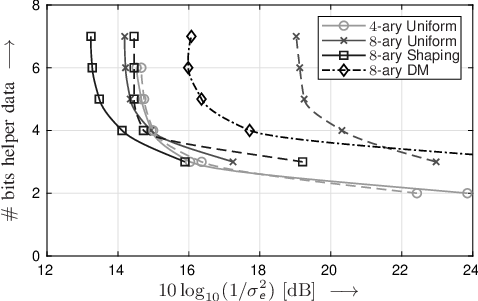}}
\caption{\label{fig_5_4}
	Number of bits helper data per PUF node over the
	signal-to-noise ratio (in dB) which is required to guarantee
	the target word error ratio of $\mathrm{WER} = 10^{-6}$.
	$4$-ary uniform, $8$-ary uniform, and $8$-ary shaping schemes with
	multilevel codes ($\lambda = 0.70$).
	Conversion helper scheme from Sec.~\ref{sec_3} and new rate design.
	Solid lines: MLC with MSD;
	Dashed lines: BICM.
	Dashed-dotted lines: distribution matcher with BICM.
	Right bottom to left top: $S = 1$, $2$, $4$, $8$, and $16$.
}
\end{figure}

Third, the tradeoff between performance and the amount of required helper data
is evaluated. The target word error ratio is set to $\mathrm{WER} = 10^{-6}$ and
the rate is still $R = 1.50~\mathrm{bit/node}$ (message length $k=1536$).
The conversion scheme requires $\log_2(M)$ bits helper data per PUF node.
The $S$-metric approach needs $\log_2(S)$ extra bits.
In Fig.~\ref{fig_5_4}, this amount of data is plotted over the signal-to-noise
ratio that is at least required to guarantee the target word error ratio.

Increasing $S$ improves performance, but also increases the number of bits
for the helper data. $S > 8$, i.e., more than $3$~extra bits, is not rewarding
as the gain in signal-to-noise ratio saturates.

The solid lines depict the results for the MLC scheme, the dashed lines when
BICM \cite{Caire:98} is used. Here, the set-partition label is replaced by a
Gray label. In each case, MLC outperforms BICM. For the $4$-ary uniform scheme
only a marginal loss is present. This is due to the fact that rate spread of
the levels (see Tab.~\ref{tab_5_1}) is not too large.
Here, BICM employs a Polar code of length $2n$ and rate $.75$. This code has to
average over the different capacities of the bit levels, resulting in a loss.
Moreover, one code of length $2n$ has to be decoded, which requires somewhat
more effort than decoding two codes of length $n$.

A different effect is visible for the $8$-ary shaping scheme. The two LSBs are
coded using a Polar code of length $2n$ and rate $.5$ (thus $n$ message bits);
the shaping level is uncoded and carries $n/2$ message bits summing up to the
desired rate. However, the spread of the level capacities is much larger
($R_0 \approx .1$ and $R_1 \approx .9$) and the code has to average over
bit channels with significantly different performance, resulting in a larger
loss.

The performance of BICM in the case of the $8$-ary uniform scheme is
significantly worse than that of the MLC scheme. Here, a Polar code of length
$4n$ with rate $3/8$ is used. By (random) puncturing $n$ codesymbols, a rate
$1/2$ code with codelength $3n$ is obtained (the puncturing pattern is, of
course, known to the decoder). Since this weak code has to average over
bit channels whose capacities have a large spread
($R_0 \approx .1$, $R_1 \approx .4$, and $R_2 \approx .9$), a poorer
performance is obtained. In addition, here a code of length $4n$ has to be
decoded, which leads to a higher numerical complexity than the MLC scheme,
where three codes of length $n$ have to be decoded.

Finally, a (constant composition) distribution matcher \cite{Schulte:16}
combined with bit-interleaved coded modulation \cite{Caire:98} as
proposed in \cite{Boecherer:15} for combined coded modulation/shaping schemes
is studied. In contrast to the multilevel coding/trellis shaping approach
\cite{Fischer:22}, here the shaping part is done first and then, using a
systematic encoder, the channel coding part.

An $8$-ary scheme with scaling $\lambda = 0.70$ in also used here.
The probabilities of the four possible amplitudes are calculated according
to (\ref{eq_3_1}). The DM generates the amplitudes according to these
probabilities. Within the block of $n=1024$ symbols, the amplitudes $1$, $3$,
$5$, and $7$ occur $530$, $330$, $128$, and $36$ times.
The binary representation of the amplitudes contains $2n = 2048$ binary
symbols. For a fair comparison, a Polar code is used as above.
Since $3n = 3072$ code bits are needed to map to $n = 1024$ amplitude
coefficients, a Polar code with a codelength of $4n = 4096$ is employed;
systematic encoding is used and $1024$ parity symbols are punctured to
obtain a rate-$2/3$ code.

The performance of this approach falls behind the multilevel coding/trellis
shaping scheme.
This is due to the fact that i) in MLC, the component codes are perfectly
matched to the actual situation, whereas BICM always has a (small) loss
because all bit are decoded in one step, as opposed to the successive
multistage decoding over the levels, taking the decoding results of the lower
levels into account (chain rule), and ii) the punctured rate $2/3$-code
appears to have worse performance than the low-rate code at level $0$ in the
MLC construction (which typically limits performance in an MLC scheme).
Also, BICM is best suited for Rayleigh fading channels, which is not the case
here.

\subsection{Evaluation with Measurement Data}

The performance of the schemes is evaluated not only by numerical simulations
but also based on measured data. To this end, $22$ instances of ROPUFs
were implemented on FPGA evaluation boards at the Institute of
Microelectronics, Ulm University. From the available ROs, $n = 1024$
disjoint pairs were randomly selected. Details can be found in
\cite{herkle2019depth}.

In the initialization phase, $10$ readouts were measured and averaged at a
temperature of $20~{}^\circ\mathrm{C}$. This average word is set as the
reference readout $\ve{x}_\mathrm{puf}$ of the respective PUF instance.
For each PUF instance the message $\ve{\ff{m}}$ is drawn at random and the
helper data is generated as described above.

In the reproduction phase, the temperature is adjusted from
$-10~{}^\circ\mathrm{C}$ in steps of $10~{}^\circ\mathrm{C}$ to
$50~{}^\circ\mathrm{C}$. For each of these temperatures, $10,\!000$ readouts
were measured per PUF instance (a total of $70,\!000$ readouts per PUF
instance). Each readout was decoded (assuming MLC and the conversion helper
data scheme) and it was recorded whether the results agreed with the selected
message or not.

\begin{figure*}[htb]
\centerline{\includegraphics[width=.80\textwidth,clip=]{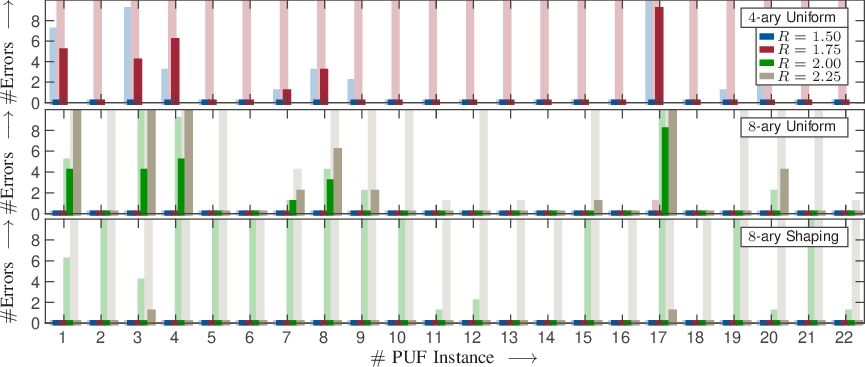}}
\caption{\label{fig_5_5}
	Number of decoding errors for $22$~PUF instances.
	Top: $4$-ary uniform scheme;
	Middle: $8$-ary uniform scheme;
	Bottom: $8$-ary shaping scheme.
	Rates and corresponding message length (bars from left to right):
	$R=1.50$ ($k=1536$), $R=1.75$ ($k=1792$), $R=2.00$ ($k=2048$),
	and $R=2.25$ ($k=2304$).
	Light bars: $S = 1$ (conventional helper data scheme);
	Dark bars: $S = 4$ ($S$-metric scheme).
}
\end{figure*}

In Fig.~\ref{fig_5_5}, the number of decoding errors is shown as a bar chart
for all $22$~PUF instances and the three discussed schemes.%
\footnote{It is assumed that at the highest bit level in the $8$-ary
	shaping scheme an additional code (with hard-decision decoding)
	is present that is able to correct up to $6$ bit errors per word
	of length $n/2$.
}
The bars correspond to the rates (from left to right) $R=1.50$ (message length
$k=1536$), $R=1.75$ ($k=1792$), $R=2.00$ ($k=2048$), and $R=2.25$ ($k=2304$).
As the $4$-ary scheme cannot support rates lager than or equal to $2.0$, the
respective bars are not shown. The light bars in the background are valid
for $S = 1$ (conventional helper data scheme) and the dark bars are valid for
$S = 4$ ($S$-metric scheme).

It is obvious that the number of errors increases as the rate increases.
For a rate of $R=1.75$, a huge number of errors occurs if the conventional
$4$-ary uniform scheme is used (the bars are clipped at $10$, but here more
than $10^5$ errors occur). Applying $S=4$ dramatically reduces the number of
errors; only $6$ PUF instances show any errors at all.

The $8$-ary schemes are error free for rates less than $R=2.00$.
The uniform scheme exhibits more errors than the shaping scheme. As before,
$S = 4$ provides a significant reduction in the number of errors.
For $R=2.25$, only two PUF instances ($\#3$ and $\#17$) show a single decoding
failure (out of $70,\!000$ words). In summary, using the proposed coding and
helper data schemes, high rates (long messages) can be retrieved very reliably.

%
%

\section{Summary and Conclusions}
\label{sec_6}

\noindent
In this paper, we have considered the generation and usage of helper data for
PUFs that provide real-valued readout symbols. A model of the readout process
as a digital transmission with randomness at the transmitter has been has been
studied.
By using coded modulation and signal shaping, a scheme is obtained that is
matched to the (approximately) Gaussian distribution of the readout.
An appropriate helper data scheme for this setting has been presented.
Compared to the literature, where a permutation has been proposed, better
performance is achieved and the amount of helper data is significantly reduced.
In addition, the generation of additional helper data, which is not necessary
for enabling decoding in the first place, but which supports the decoding
process and increases reliability, has been discussed.

By means of numerical simulations and the evaluation of measurement data,
it has been shown that the $8$-ary scheme with multilevel coding and trellis
shaping shows the best performance. As long as the readout process exhibits
a signal-to-noise ratio greater than about $13~\dB$, the rate per PUF node
that can be reliably extracted can be as high as $2~\mathrm{bit/node}$.

In this paper, the numerical examples have been given for a ring-oscillator
PUF as a representative of so-called silicon PUFs. However, the approaches
can be applied to any type of PUF, provided that real-valued readout symbols
are delivered. The code design can be easily adapted to other distributions
than the Gaussian one.

An interesting field for the application of the discussed schemes are so-called
\emph{channel PUFs}, where two communication partners agree on a secret key
based on jointly available channel measurements, see, e.g., the surveys in
\cite{Wang:15,Zhang:16} and the references therein. Very different approaches
are available, e.g., \cite{Hershey:95,Tope:01,Ye:06,Azimi:07} to name
only a few. Often only the received signal strength is used, e.g.,
\cite{Jana:09,Premnath:13}, or the agreement of the keys of both partners
is ensured by information reconciliation schemes on the protocol level, see,
e.g., the overview in \cite{Huth:16}.

The channel coefficients in a broadband, frequency-selective channel are
typically Gaussian distributed. If the channel is observed at frequencies
spaced (at least) by the \emph{coherence bandwidth}, the coefficients can
be assumed to be drawn independently. The channel coefficients at the
different frequencies are thus the PUF nodes and the set of all measured
channel coefficients gives the PUF. The coded modulation/shaping schemes
in combination with the helper data schemes discussed in this paper are
well suited for use in channel PUFs.
One communication partner carries out the steps of the initialization phase,
i.e., it randomly draws the message and, knowing the channel measurements,
generates the helper data. The helper data can be transmitted publicly to
the other communication partner. It carries out the steps of the reconstruction
phase, i.e., knowing its channel measurements (which differ slightly from
those of the other partner) and the helper data, it decodes the message.
If the signal-to-noise ratio of the channel measurements is large enough,
long keys can be agreed with high reliability.
The details are subject of ongoing work.

%
%

\appendices
\section{Transformation of Gaussians}
\label{sec_a}

\noindent
In order to work more conveniently with the Gaussian PUF readout and
regions that have been adjusted to uniform probabilities, we introduce a
handy transformation and its inverse.

Let the transformation be given by
\begin{eqnarray}						\label{eq_a_}
\tilde{x} \;=\; g(x) &\defeq&
			\erf\Big( \tfrac{x}{\sqrt{2}} \Big) \;,
\\[2mm]
\intertextsl{where}						\label{eq_a_2}
\erf(z) &\defeq& \frac{2}{\sqrt{\pi}} \int_{0}^z \e^{-t^2} \;\d t
\end{eqnarray}
is the error function. Note that the inverse transformation is given by
\begin{eqnarray}						\label{eq_a_3}
x \;=\, g^{-1}(\tilde{x}) &=&
		\sqrt{2}\erf^{-1}( \tilde{x} ) \;.
\end{eqnarray}
Subsequently, all quantities in the \emph{transform domain} are marked with a
tilde (``tilde domain'').

Let $\rv{x}$ be a Gaussian random variable with zero mean and unit variance
(e.g., the readout). Its pdf reads
\begin{eqnarray}						\label{eq_a_4}
\pdf_\rv{x}(x) &=& \frac{1}{\sqrt{2\pi}}\, \e^{-\frac{x^2}{2}} \;.
\end{eqnarray}
It is straightforward to show (e.g., \cite{Papoulis:02}) that $g(x)$
transforms the Gaussian random variable $\rv{x}$ into the random variable
$\tilde{\rv{x}}$, which is uniformly distributed over the interval
$[-1,\; 1]$, i.e.,
\begin{eqnarray}						\label{eq_a_5}
\pdf_{\tilde{\rv{x}}}(\tilde{x}) &=&
		\cases{\frac{1}{2} \;, & $-1 \le \tilde{x} \le 1$ \cr
			0 \;, & else} \;.
\end{eqnarray}

For $M$-ary uniform signaling, the limits $L_\rho$ have to be chosen so that
\begin{eqnarray}						\label{eq_a_6}
\int_{-\infty}^{L_\rho} \frac{1}{\sqrt{2\pi}}\,\e^{-\frac{t^2}{2}} \;\d t
	&=& \frac{\rho}{M} \;,\qquad \rho=0,\ldots,M \;.
\end{eqnarray}
Via $g(x)$, these limits are transformed to uniformly spaced limits in the
transform domain
\begin{eqnarray}						\label{eq_a_7}
\tilde{L}_\rho &\defeq& g(L_\rho) \;=\; 2\frac{\rho}{M} - 1
					\;,\qquad \rho=0,\ldots,M \;.
\end{eqnarray}
This means that the regions $\mathcal{R}_\rho$ are transformed to regions
$\tilde{\mathcal{R}}_\rho$ which all have the same width $\frac{2}{M}$.

\begin{figure}[htb]
\centerline{\includegraphics[width=.98\columnwidth,clip=]{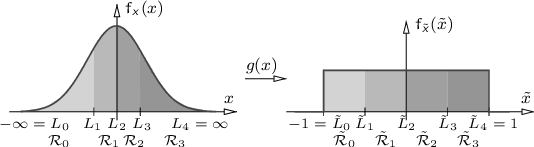}}
\caption{\label{fig_a_1}
	Transformation of a Gaussian distribution onto a uniform one and
	transformation of the limits and regions. Uniform probabilities of the
	regions. $M = 4$.
}
\end{figure}

The procedure of transforming the pdf, the limits, and the regions is
visualized in Fig.~\ref{fig_a_1} for $M = 4$.

%

\section*{Acknowledgment}

\noindent
The author would like to thank Holger Mandry and Maurits Ortmanns for
the discussions on PUFs and for providing their measurement data on ROPUFs.

%

%

\end{document}